\documentclass[a4paper,10pt,twoside]{cpc-hepnp}

\usepackage{multicol}
\usepackage{graphicx}
\usepackage{booktabs}
\usepackage{amssymb,bm,mathrsfs,bbm,amscd}
\usepackage[tbtags]{amsmath}
\usepackage{textcomp}
\usepackage{lastpage}

\begin{document}

\fancyhead[co]{\footnotesize  et al: }

\footnotetext[0]{Received 2011}


\title{Luminosity determination for the $pd$ reaction at 2.14~GeV with
  WASA-at-COSY\thanks{Supported by the Forschungszentrum J\"{u}lich
    including the JCHP-FFE program, the European Community under the
    FP6 program (Hadron Physics, RII3-CT-2004-506078) \& the FP7
    program (FP7-INFRASTRUCTURES-2008-1, Grant N.227431), the German
    BMBF, the China-German CSC-HGF exchange program, the German
    Research Foundation(DFG) and the National Natural Science Foundation
    of China (10635080,10925526)}}

\author{%
ZHENG Chuan$^{1,2,4,5;1)}$\email{zhengchuan@fudan.edu.cn} 
\quad M.~B\"uscher$^{1,2}$             
\quad P.~Fedorets$^{3}$                
\quad V.~Hejny$^{1,2}$                 
\quad H.~Str\"oher$^{1,2}$  \\         
\quad XU Hu-Shan$^{4}$         
\quad YUAN Xiao-Hua$^{1,2,4}$  
}

\maketitle

\address{%
 1~(Institut f\"ur Kernphysik, Forschungszentrum J\"ulich,
    52425 J\"ulich, Germany
    )\\
 2~(J\"ulich Center for Hadron Physics,
    Forschungszentrum J\"ulich,
    52425 J\"ulich, Germany
    )\\
 3~(Institute for Theoretical and Experimental Physics,
    State Scientific Center of the Russian Federation,
    Bolshaya Cheremushkinskaya~25,
    117218 Moscow, Russia
    )\\
 4~(Institute of Modern Physics,
    Chinese Academy of Sciences,
    Nanchang Rd.~509, 730000 Lanzhou, China
    )\\
 5~(Institute of Modern Physics,
    Fudan University,
    Handan Rd.~220, 200433 Shanghai, China
    )
}

\begin{abstract}
  The luminosity for a WASA-at-COSY experiment involving the $pd$
  reaction at 2.14~GeV proton-beam energy is determined
  by the forward $pd$ elastic scattering, which yields an average
  beam-on-target value of $[5.2 \pm 0.3({\mathrm{stat}}) \pm 0.3({\mathrm{syst}})]
  \times 10^{30}\,\mathrm{s}^{-1} \mathrm{cm}^{-2}$. In addition,
  the forward $pd$ elastic-scattering angular distribution is
  obtained with four-momentum transfer squared $-t$ between 0.16~(GeV/c)$^{2}$
  and 0.78~(GeV/c)$^{2}$ at this beam energy, which is compared with
  other experimental data and the $pd$ double scattering model.
\end{abstract}




\begin{multicols}{2}

\section{Introduction}

The WASA detector facility~\cite{Barg08}, a nearly $4\pi$
multidetector system equipped with an internal $\rm{H_{2}/D_{2}}$
frozen-pellet target~\cite{Eks02}, is now operated at the COoler
SYnchrotron COSY-J\"ulich~\cite{Mai97} which delivers proton and
deuteron beams with momenta up to 3.7 GeV/c. It serves for the
investigation of hadronic processes and systems, such as light meson
production, exotic hadronic states, symmetries and their breaking,
following $pp$, $pd$ and $dd$ reactions~\cite{WASA04}. WASA is well
suited to detect both charged and neutral decays, {\em e.g.\/} of the light scalar
mesons $a_{0}/f_{0}(980)$ produced in such reactions. A first test
measurement has been performed in order to obtain the yet unknown
light-scalar production cross sections in $pd\to{}^{3}A\,X$
processes at a beam energy of $T_{p}=2.14$ GeV from the measurement
of its strong decays. We describe here the luminosity determination
for this experiment.

The luminosity is defined as the number of beam particles passing
through the target per unit time multiplied by the number of atoms
in the target per unit area. In the case of a stored beam and an
internal pellet target, the luminosity can be inferred from the
number of beam particles, the beam-revolution frequency, the number
of atoms in one pellet, and the beam-pellet overlap factor. The
parameters of the COSY beam and the pellet target for this
particular experiment are listed in Table~\ref{tab1}.

The proton beam in the 184~m long storage ring, circulating with a
revolution frequency $f$ related to its energy $T_{p}$, has an
average beam intensity $N_{\mathrm{C}}$ and a
diameter~$2R_{\mathrm{C}}$ at the interaction region. The number of beam particles per unit
time and area, called current density, is obtained from the formula:
\begin{eqnarray}
  \label{eq1} j & = & (N_{\mathrm{C}} f)/({\pi}R_{\mathrm{C}}^{2}) \ .
\end{eqnarray}
On the other hand the number of atoms in one D$_{2}$-pellet can be
calculated from its radius $R_{\mathrm{t}}$ and density $\rho$ with:
\begin{eqnarray}
  \label{eq2} N_{\mathrm{t}}(\mathrm{D}) & = & \frac{4}{3}{\pi}R_{\mathrm{t}}^{3}
  {\rho}\frac{\mathrm{n}}{\mathrm{M}}{N_{\mathrm{A}}} \ ,
\end{eqnarray}
where $\mathrm{n}=2$ atoms/molecule for Deuterium,
$\mathrm{M}\approx4$ g/mol is its molar mass, and
$N_{\mathrm{A}}=6.022\times10^{23}$ molecules/mol is the Avogadro
constant. Inserting the numbers from Table~\ref{tab1}, the average
luminosity for this experiment is estimated as $L\sim 5\times
10^{30}\, \mathrm{s}^{-1} \mathrm{cm}^{-2}$ which should be regarded
as an order-of-magnitude estimate, since the beam diameter and the
pellet rate vary as a function of time.

\begin{center}
  \tabcaption{ \label{tab1} Beam and target parameters.}
  \footnotesize
  \begin{tabular*}{80mm}{lcl}
    \toprule Parameters  &  Values  & Units/remarks \\
    \hline Beam intensity($N_{\mathrm{C}}$)  &  $1.7\times 10^{9}$ & particles \\
    Revolution frequency ($f$) &  1.5564~MHz &  2.14 GeV \\
    Beam size ($2R_{\mathrm{C}}$)  &  6~mm  & 3--6~mm \\
    Current density ($j$) & $9.3\times 10^{15}$ & particles/($\mathrm{s}\cdot\mathrm{cm}^{2}$)\\

    \hline Pellet diameter (2$R_{\mathrm{t}}$) &  35~$\mu$m  & 25--35~$\mu$m\\
    D$_{2}$ density ($\rho$)  &  0.162 g/cm$^{3}$ & triple point\\
    Atoms/pellet ($N_{\mathrm{t}}$(D))  &  $1.10\times 10^{15}$ & D$_{2}$\\
    \hline
    Pellet frequency  &  6~kHz  &  5--12~kHz \\
    Pellet velocity   & 70~m/s  &  60--80~m/s  \\
    Pellets in beam   & 0.5 &  overlap factor \\
    \bottomrule
  \end{tabular*}
\end{center}

\section{Identification of $pd$ elastic scattering}

Besides the estimate outlined above, it is useful for luminosity
determination to record a monitoring reaction during the experiment,
such as elastic scattering for which the corresponding cross
sections are well known in dependence of both the beam energy and
the scattering angle. In the data acquisition (DAQ) the trigger
settings for two charged tracks within WASA were applied to select
elastic-scattering events. Both $pd$ elastic scattering and $pp$
quasi-elastic scattering, with the neutron being a spectator
particle, could be recorded by the same trigger condition.

For the scattering processes the data analysis starts with the
selection of one charged track in the forward detector (FD, covering
emission angles of $\theta = 3 - 18$\textdegree) and the second one
in the central detector (CD, $\theta = 20-170$\textdegree). After a
FD-CD time correlation cut, the coplanarity of the two charged
tracks is exploited. The corresponding azimuthal-angle correlation
is plotted in Fig.~\ref{fig1}. For the azimuthal-angle correlated
events, the polar-angle correlation of the charged FD and CD tracks
is also depicted in Fig.~\ref{fig1}. It is seen that most events are
located around the $pp$ elastic line (determined from a MC
simulation) and thus stem from $pp$ quasi-elastic scattering. The MC
simulations are carried out with the Pluto event
generator~\cite{Pluto} and the WASA detector simulation package
based on GEANT3. The peak around 180\textdegree\ of the
azimuthal-angle correlation has a width of $\sigma =
6.9$\textdegree\ in data, which is reproduced by the simulation of
$pp$ quasi-elastic scattering, while for $pp$ elastic scattering a
much smaller width of $\sigma = 2.0$\textdegree\ is obtained.

\begin{center}
  \includegraphics[width=0.3\textwidth]{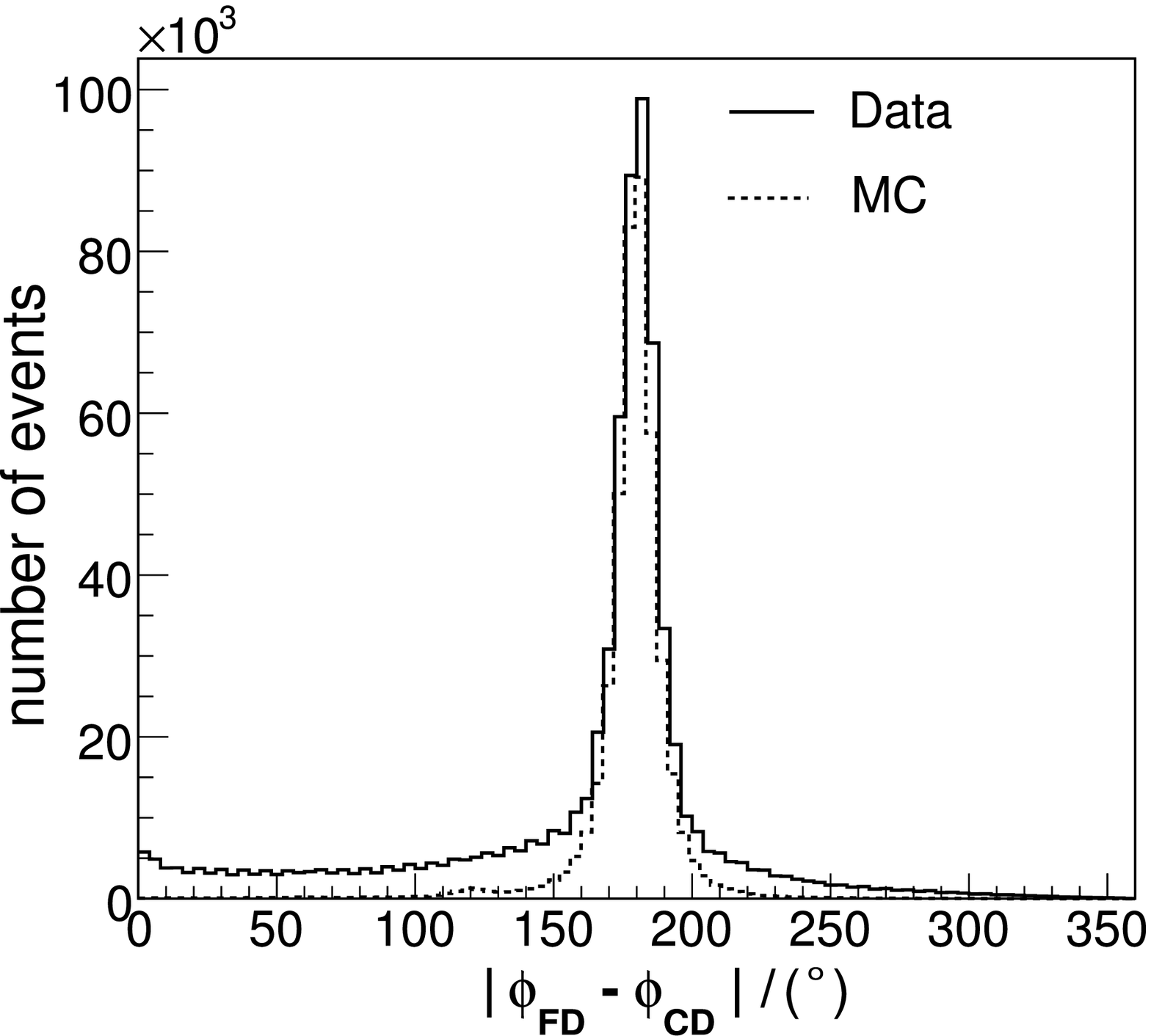}\\
  \vspace*{5mm}
  \includegraphics[width=0.3\textwidth]{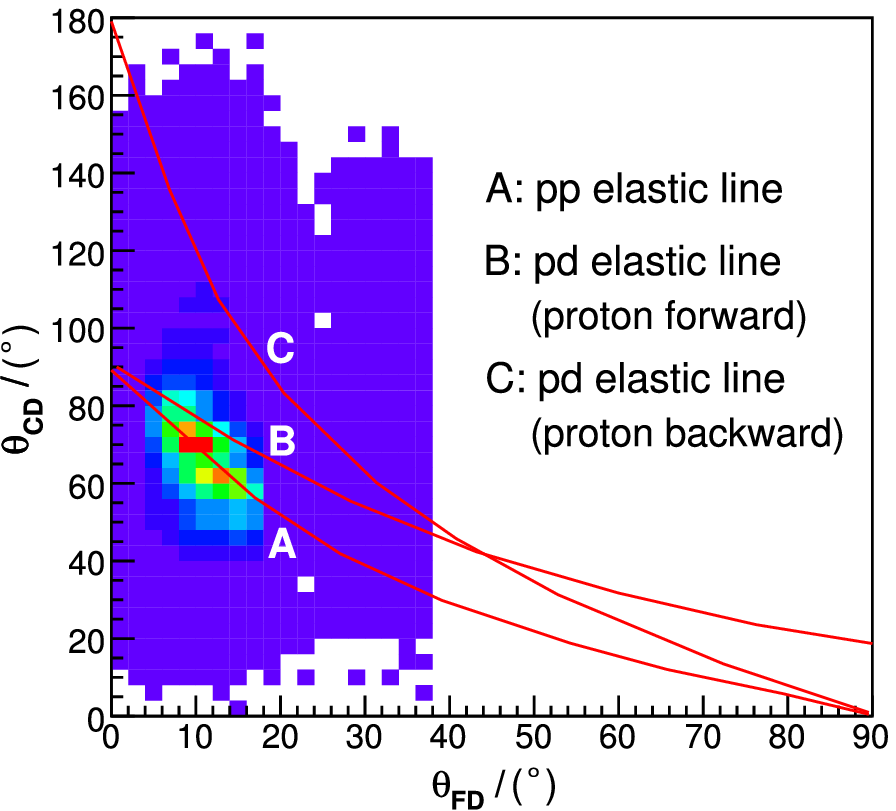}\\
  \figcaption{\label{fig1} Angular correlation of two charged tracks
    for the azimuthal (upper panel) and polar (lower) angles.
    The dashed line presents the MC simulation of $pp$ quasi-elastic scattering from $pd$ reaction.
    Line A depicts the kinematics of $pp$ elastic scattering;
    Lines B and C are for $pd$ elastic scattering with the proton emitted in forward and backward direction, respectively.}
\end{center}

The $pd$ elastic-scattering events, with the proton detected in the
FD and the deuteron in the CD, can be identified on top of a huge
background from the above mentioned $pp$ quasi-elastic scattering.
For the charged track in the CD, the $\Delta \mathrm{\it{E}}$-{\em
vs.}-momentum distribution is presented in Fig.~\ref{fig2} together
with the expected proton and deuteron lines obtained from a MC
simulation. A weak cut is applied, which then allows one to extract
the $pd$ elastic-scattering events from the strongly reduced $pp$
quasi-elastic background. For the deuteron candidates the agreement
between the polar-angle value measured directly in the CD and the
one calculated from the other charged track in the FD is checked on
an event-by-event basis. Figure~\ref{fig2} shows that a clear peak
of forward $pd$ elastic scattering appears around zero angular
difference. We did not further analyze the backward $pd$ elastic
scattering with the deuteron in the FD and the proton in the CD
since there are very few events around Line C shown in
Fig.~\ref{fig1}.

\begin{center}
  \includegraphics[width=0.3\textwidth]{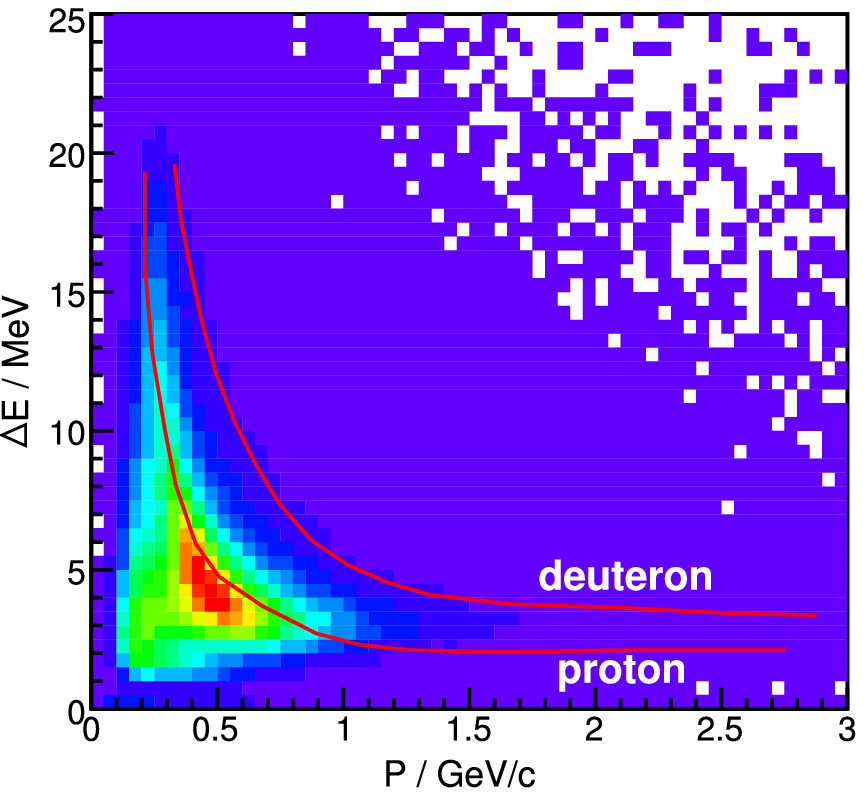} \\
  \vspace*{5mm}
  \includegraphics[width=0.3\textwidth]{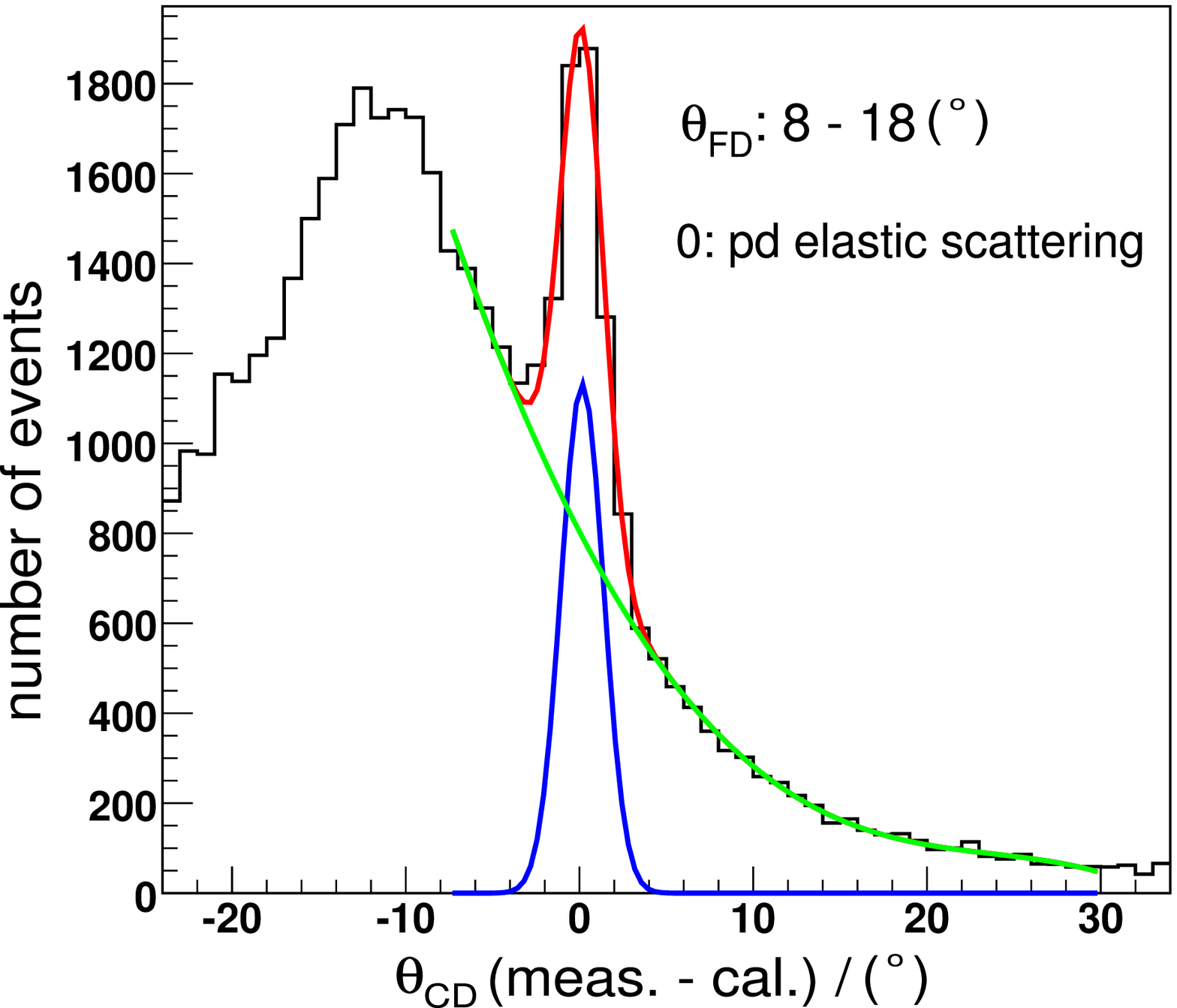} \\
  \figcaption{\label{fig2} Upper panel: Energy deposited in the plastic
    scintillator {\em vs.} momentum reconstructed in the solenoid
    field of WASA. Lower panel: Difference of the measured polar angle of
    the charged track in the CD and the one calculated from the
    measured polar angle in the FD assuming forward $pd$ elastic
    kinematics. The background under the $pd$ elastic-scattering peak
    has been fitted by a fourth order polynomial and is shown by the green line (color online).
    The remaining signal is also indicated in the figure by a Gaussian distribution. }
\end{center}

\section{Normalization of $pd$ angular distribution}

The normalization of $pd$ elastic-scattering angular distribution
$(d{\mathrm{N}}/d\Omega)_{\mathrm{lab}}$ to the differential cross
sections $(d\sigma/d\Omega)_{\mathrm{lab}}$ is described by the
formula:
\begin{eqnarray}
  \label{eq3} L_{\mathrm{int}} & = & (\frac{d{\mathrm{N}}}{d\Omega})_{\mathrm{lab}} /
  (\frac{d\sigma}{d\Omega})_{\mathrm{lab}} \ ,
\end{eqnarray}
where $L_{\mathrm{int}}$ is the integral luminosity for the
corresponding beam time.

The forward $pd$ elastic-scattering angular distribution from our
data with four-momentum transfer squared $-t$, given in
Table~\ref{tab2}, is obtained in a polar-angle interval of
1\textdegree\ in the FD with the $pd$ elastic-scattering peak
observed starting from 8\textdegree\ to 18\textdegree\ in the lab
frame. The $-t$ is the Lorentz-invariant momentum transfer ranging
from 0.16~(GeV/c)$^{2}$ to 0.78~(GeV/c)$^{2}$. The number of $pd$
elastic-scattering events is extracted from the number of total
events in the gaussian ${\pm}3\sigma$ region with the relative area
ratio of the gaussian peak to the polynomial background, shown in
Fig.~\ref{fig2}, and the statistical error is taken from the number
of total events and the fitting error of the polynomial background.
The geometric acceptance, reconstruction and cut efficiencies have
been determined as a whole from the MC simulation for each
polar-angle interval as listed in Table~\ref{tab2}. In addition, the
trigger pre-scaling factor is 2~000 for the first 35 runs and then
4~000 for the later 123 runs. The DAQ lifetime correction is
estimated with a relative error 12.4\%. Both, the statistical error
and the DAQ lifetime-correction error, are listed in
Table~\ref{tab2}.

\begin{center}
\tabcaption{ \label{tab2}  Forward $pd$ elastic-scattering angular
distribution at 2.14~GeV from our data. The last column
$(d{\mathrm{N}}/d\Omega)_{\mathrm{lab}}$ includes the statistical
error of the $pd$ elastic-scattering events and the error of the DAQ
lifetime correction.} \footnotesize
\begin{tabular*}{80mm}{cccc}
\toprule proton, $\theta_{\mathrm{lab}}$ &  $-t$           & efficiency    & $(d{\mathrm{N}}/d\Omega)_{\mathrm{lab}}$ \\
            (\textdegree)                &  (GeV/c)$^{2}$  & (\%)          & ($\times 10^{6}$/sr)    \\  \hline
               8.5                       &  0.186          & 10.4          & 5640$\pm$190$\pm$699   \\
               9.5                       &  0.231          & 42.0          &  884$\pm$31 $\pm$110    \\
               10.5                      &  0.281          & 49.3          &  532$\pm$24 $\pm$66    \\
               11.5                      &  0.336          & 51.1          &  314$\pm$16 $\pm$39    \\
               12.5                      &  0.394          & 51.2          &  238$\pm$12 $\pm$30    \\
               13.5                      &  0.457          & 51.6          &  175$\pm$9  $\pm$22    \\
               14.5                      &  0.523          & 52.2          &  210$\pm$11 $\pm$26    \\
               15.5                      &  0.594          & 52.4          &  139$\pm$8  $\pm$17   \\
               16.5                      &  0.668          & 50.2          &  138$\pm$8  $\pm$17   \\
               17.5                      &  0.745          & 38.2          &  118$\pm$10 $\pm$15    \\
\bottomrule
\end{tabular*}
\end{center}

The forward $pd$ elastic-scattering differential cross sections from
2.0~GeV data~\cite{Col67} with four-momentum transfer squared $-t$
between 0.35~(GeV/c)$^{2}$ and 1.0~(GeV/c)$^{2}$ are fitted with the
exponential function $a\cdot\mathrm{e}^{b(-t)}$, shown in
Fig.~\ref{fig3}, which has a slope value $b_{1}=-1.87\pm 0.09$,
while a slope $b_{2}=-1.84\pm 0.50$ is obtained from fitting our
data independently. When the slope value is fixed as $b=-1.87$ for
both cases, the fitting procedure gives the coefficient
$a_{1}=710\pm 11~\mu$b/sr for 2.0~GeV data and $\tilde{a}_{2}=465\pm
26~\mu$b/sr$\cdot$pb$^{-1}$ for our data, respectively. Then, a
scaling factor is applied to normalize our data to the same height
of 2.0~GeV data, which leads to the integral luminosity.

\begin{center}
  \includegraphics[width=0.35\textwidth]{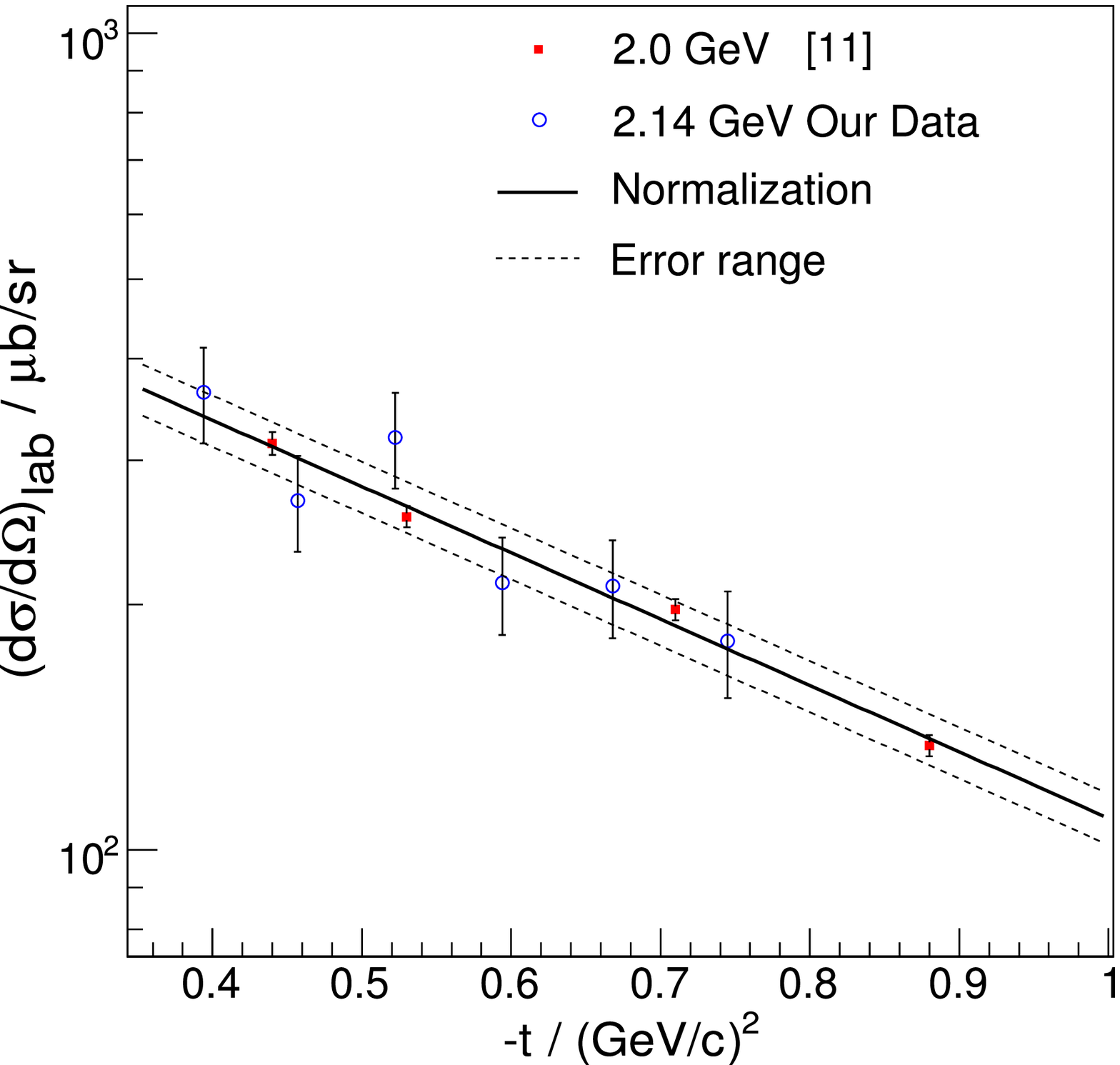}\\
  \figcaption{\label{fig3} Normalization of the $pd$
    elastic-scattering angular distribution from our data to the differential cross sections from 2.0~GeV data
    with $-t$ between 0.35~(GeV/c)$^{2}$ and 1.0~(GeV/c)$^{2}$. The fit function is an
    exponential.}
\end{center}

The $pd$ elastic-scattering differential cross sections for protons
in the forward direction with four-momentum transfer $-t$ less than
1.0~(GeV/c)$^{2}$ are plotted in Fig.~\ref{fig4} for proton-beam
energies ranging from 0.425~GeV to
11.90~GeV~\cite{Boo71,Bos72,Guel91,Win80,Col67,Bra70}.

\begin{center}
  \includegraphics[width=0.35\textwidth]{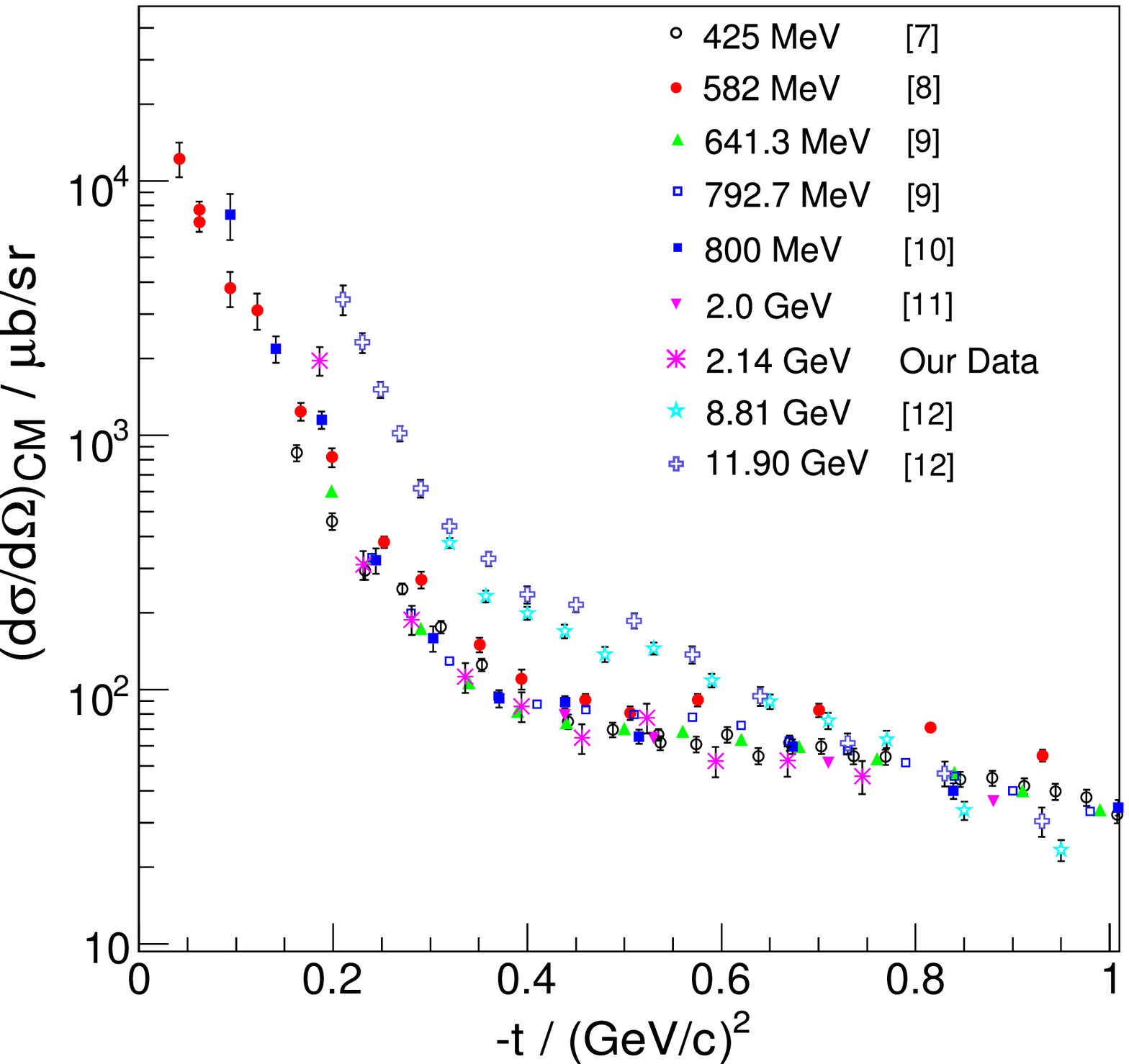}\\
  \figcaption{\label{fig4} Compilation of forward $pd$
    elastic-scattering differential cross sections {\em vs.\/}
    four-momentum transfer $-t$ including our data after the
    normalization.}
\end{center}

The forward $pd$ elastic-scattering angular distribution can be
explained by nucleon-nucleon single and double scattering and their
interference for the deuteron as a double scatterer
~\cite{Gla66,Fra66}. A further theoretical investigation, which
takes the $d$-state admixture fully into account~\cite{Fra69}, gives
a better description of the shoulder near $-t \approx
0.44$~(GeV/c)$^{2}$. Our data points are in good agreement with the
model calculations around this sensitive region (see
Fig.~\ref{fig5}) which provides further confidence in the
normalization procedure.

\begin{center}
  \includegraphics[width=0.35\textwidth]{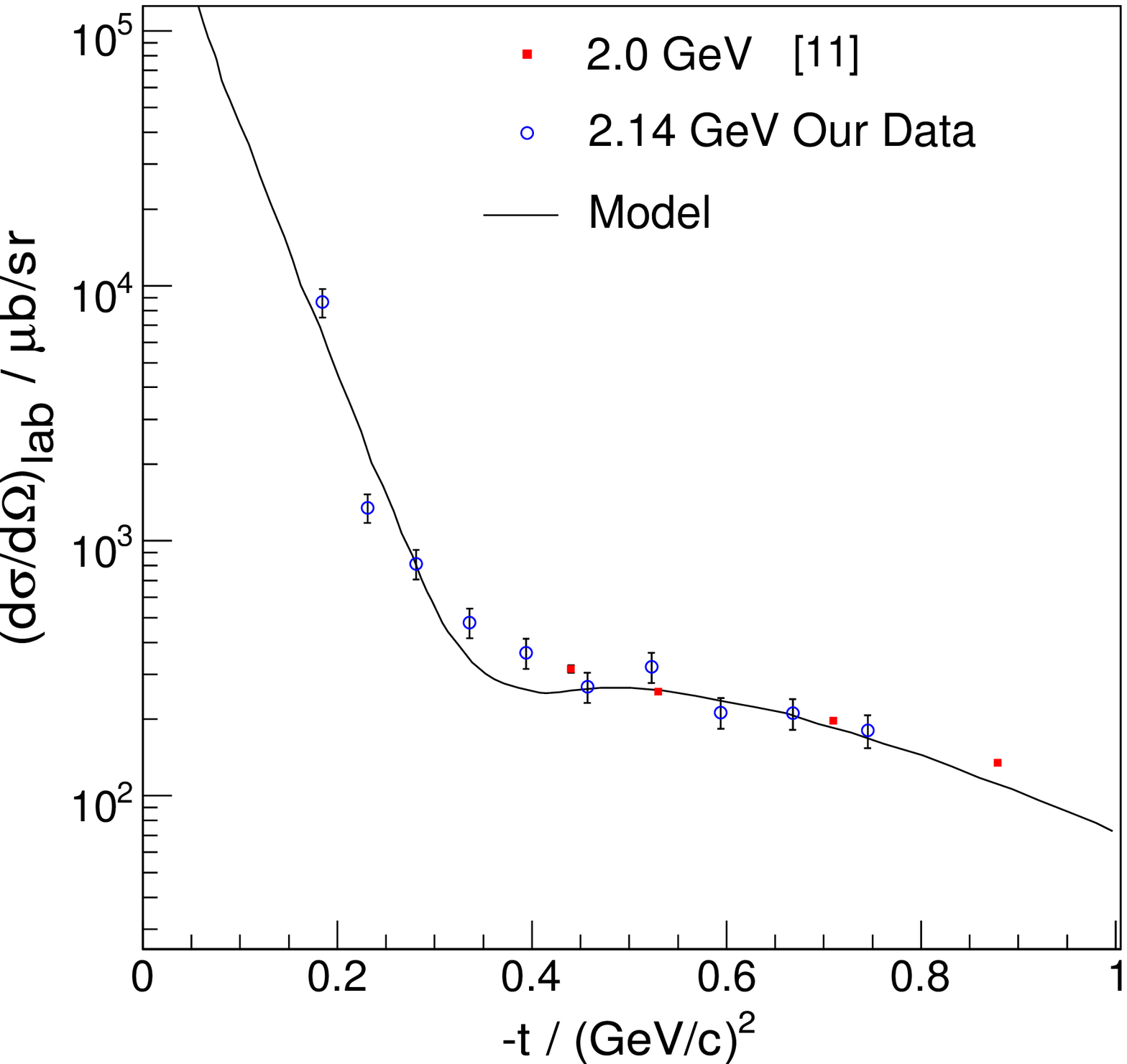}\\
  \figcaption{\label{fig5} Comparison of the forward $pd$ elastic-scattering angular distribution
  available from our data with the model calculation~\cite{Fra69} after the normalization.}
\end{center}

\section{Result and error estimate}

The average luminosity in the units of ($\mathrm{s}^{-1}
\mathrm{cm}^{-2}$) can be obtained from the integral luminosity
divided by the effective beam time, which discards the rest time
between the cycles. There are 274 runs in this experiment with a
data-taking time of about 80 hours, of which the first 158 runs have
the valid trigger condition for the forward $pd$ elastic scattering
with the effective beam time $1.19\times 10^{5}$ seconds. The
average luminosity is obtained from this part of the beam time.

In the normalization of the angular distribution, the beam energy
difference between our data and 2.0~GeV data is not considered,
which could introduce a small correction. The $pd$
elastic-scattering differential cross sections in the lab frame with
four-momentum transfer squared $-t$ at 0.44~(GeV/c)$^{2}$ are
plotted in Fig.~\ref{fig6} for proton-beam energies ranging from
0.425~GeV to 11.90~GeV~\cite{Boo71,Guel91,Col67,Bra70}. After
comparing the agreement between data and three kinds of fit
functions, the quadratic one is used to estimate the correction
factor with a value $c=1.06\pm 0.07$ for the differential cross
section when the beam energy increasing from 2.0~GeV to 2.14~GeV.

\begin{center}
  \includegraphics[width=0.35\textwidth]{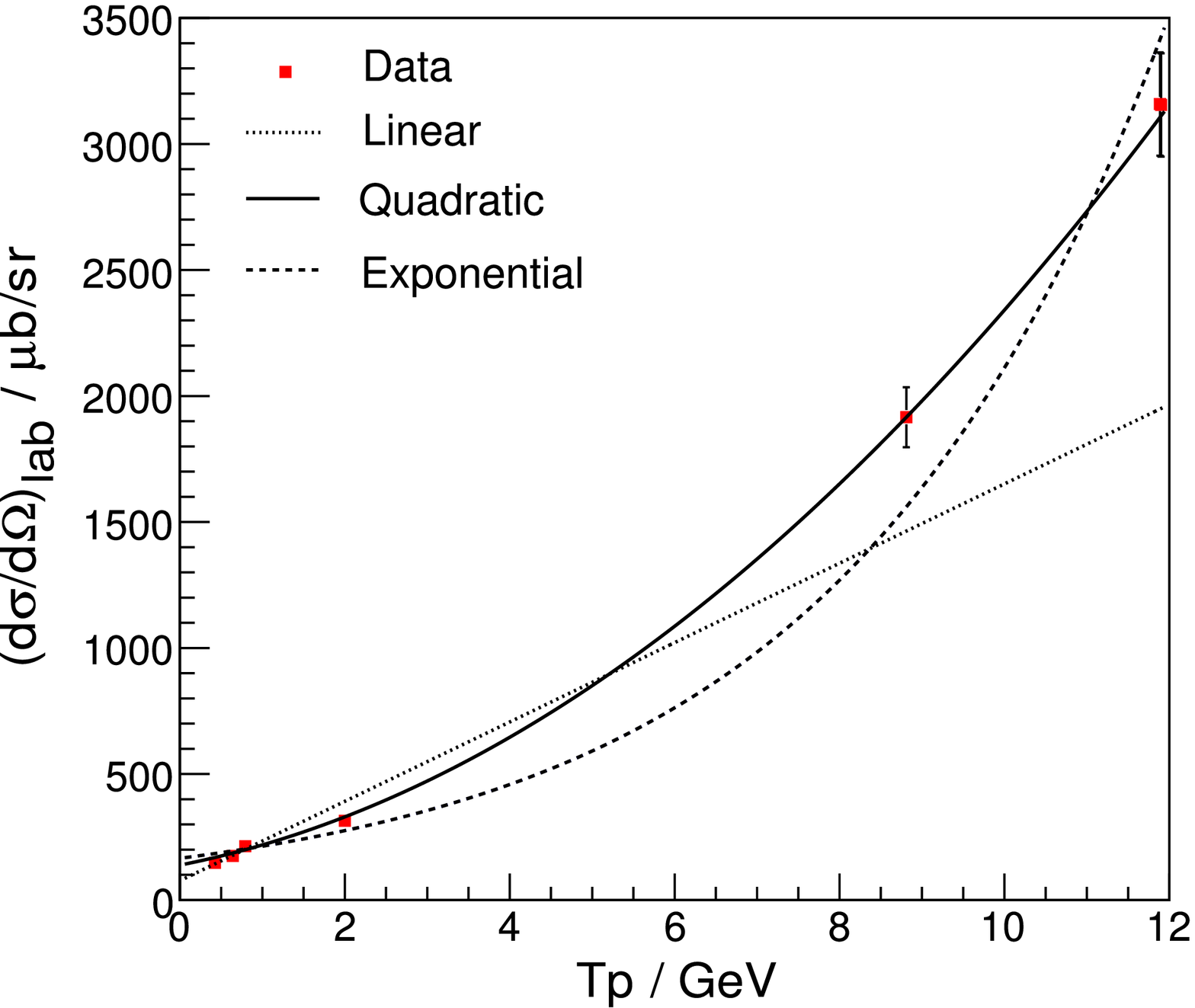} \\
  \figcaption{\label{fig6} Fit of $pd$ differential cross
    sections at the same $-t$ value {\em vs.\/} proton-beam energy.
    Three kinds of fit functions are compared.}
\end{center}

Then, the integral luminosity is deduced as
\begin{eqnarray}
  \label{eq4} L_{\mathrm{int}}(158~\rm{runs}) \,=\, \frac{\tilde{a}_{2}}{c\cdot a_{1}} \nonumber \\
    \,=\, 0.62 \pm 0.03({\mathrm{stat}}) \pm 0.04({\mathrm{syst}})~{\mathrm{pb}}^{-1} \ ,
\end{eqnarray}
with the statistical error from $\tilde{a}_{2}$ and the systematic
error from $c$ and $a_{1}$. This yields the average
luminosity during beam-on-target times as
\begin{eqnarray}
  \label{eq5} L \,=\, [5.2 \pm 0.3({\mathrm{stat}}) \pm 0.3({\mathrm{syst}})]
  \times 10^{30}\,\mathrm{s}^{-1} \mathrm{cm}^{-2} \ .
\end{eqnarray}

Finally, it is possible to estimate the total integral luminosity
with the help of another monitoring trigger defined as two charged
tracks in the central detector, which was available during the whole
beam time. These two monitoring triggers have almost the same rates
after pre-scaling, and their ratio is more or less stable within a
relative error of 6.5\%. Including the above relative error 12.4\%
from the DAQ lifetime correction, the ratio of the total integral
luminosity to the partial integral luminosity of Eq.~\ref{eq4} is
$1.54\pm 0.22$, which yields
\begin{eqnarray}
  \label{eq6} L_{\mathrm{int}}(274~\rm{runs})  \,=\,  \nonumber \\
   0.95 \pm 0.05({\mathrm{stat}}) \pm 0.15({\mathrm{syst}})~{\mathrm{pb}}^{-1} \ .
\end{eqnarray}

\section{Summary}

The aim of this experiment is to measure the light scalar meson
$a_{0}/f_{0}(980)$ production in $pd\to{}^{3}A\,X$ reactions, while
the $pd$ elastic scattering is measured in parallel as a reference
reaction to determine the luminosity. As the effective beam time is
the same for both aim and reference reactions, the value of the
integral luminosity in Eq.~\ref{eq6} is used for the evaluation of
the $a_{0}/f_{0}(980)$ production cross sections in $pd$ reactions.

\bigskip

\acknowledgments{We gratefully thank the COSY operators, the
technical and administrative staff at the Forschungszentrum
J\"{u}lich as well as all the members of the WASA-at-COSY
collaboration for their support.}

\end{multicols}

\vspace{-2mm}
\centerline{\rule{80mm}{0.1pt}}
\vspace{2mm}

\begin{multicols}{2}

\end{multicols}

\vspace{5mm}

\clearpage


\begin{thebibliography}{90}

\vspace{3mm}

\bibitem{Barg08} Bargholtz C. et al. Nucl. Instr. Meth. A, 2008, {\bf 594}: 339---350

\bibitem{Eks02} Ekstr\"{o}m C. and CELSIUS/WASA Collaboration. Phys. Scripta, 2002, {\bf T99}: 169---172

\bibitem{Mai97} Maier R. Nucl. Instr. Meth. A, 1997, {\bf 390}: 1---8

\bibitem{WASA04} WASA-at-COSY Collaboration. arXiv: nucl-ex/0411038


\bibitem{Roh04} Rohdje{\ss} H. et al. arXiv: nucl-ex/0403043

\bibitem{Pluto} Fr\"{o}hlich I. et al. PoS ACAT2007: 076; arXiv: nucl-ex/0708.2382

\bibitem{Boo71} Booth N.E. et al. Phys. Rev. D, 1971, {\bf 4}: 1261---1267

\bibitem{Bos72} Boschitz E.T. et al. Phys. Rev. C, 1972, {\bf 6}: 457---466

\bibitem{Guel91} G\"{u}lmez E. et al. Phys. Rev. C, 1991, {\bf 43}: 2067---2076

\bibitem{Win80} Winkelmann E. et al. Phys. Rev. C, 1980, {\bf 21}: 2535---2541

\bibitem{Col67} Coleman E. et al. Phys. Rev., 1967, {\bf 164}: 1655---1661

\bibitem{Bra70} Bradamante F. et al. Phys. Lett. B, 1970, {\bf 32}: 303---308

\bibitem{Fra69} Franco V. and Glauber R.J. Phys. Rev. Lett., 1969, {\bf 22}: 370---374

\bibitem{Gla66} Franco V. and Glauber R.J. Phys. Rev., 1966, {\bf 142}: 1195---1214

\bibitem{Fra66} Franco V. and Coleman E. Phys. Rev. Lett., 1966, {\bf 17}: 827---830


\end{thebibliography}
\end{document}